%% file: root.tex
\documentclass[letterpaper, 10 pt, conference]{ieeeconf}  

\IEEEoverridecommandlockouts                              

\usepackage{graphics} 
\usepackage{epsfig} 
\usepackage{times} 
\usepackage{amsmath} 
\usepackage{amssymb}  
\let\labelindent\relax
\usepackage[inline]{enumitem}

\usepackage{pgf}
\usepackage{pgfplots}
\pgfplotsset{compat=newest}

\usepackage{fancyhdr}
\newcommand{\mytitle}{\textbf{Accepted version.} To appear in the proceedings of the \textit{61st IEEE Conference on Decision and Control}.\\
\copyright 2022 IEEE. Personal use of this material is permitted. Permission from IEEE must be obtained for all other uses, in any current or future media, including reprinting/republishing this material for advertising or promotional purposes, creating new collective works, for resale or redistribution to servers or lists, or reuse of any copyrighted component of this work in other works.}
\usepackage[isbn=false,
 backend=biber,
 style=ieee,
 bibstyle=ieee, 
 citestyle=numeric-comp, 
 sortcites=true, 
 maxbibnames=10,
 maxcitenames=2,
 mincitenames=1,
 giveninits=true, 
 backref=false]{biblatex}
\DeclareRobustCommand{\citet}[1]{\citeauthor{#1}~\cite{#1}}

\graphicspath{{figures/}}

\usepackage{hyperref}
\hypersetup{
    colorlinks=false,
    pdfborder={0 0 0},
    pdfborderstyle={/S/U/W 0},
}

\usepackage{nicefrac}
\usepackage{algorithm}
\usepackage{algpseudocode}
\algdef{SE}[DOWHILE]{Do}{doWhile}{\algorithmicdo}[1]{\algorithmicwhile\ #1}%

\newtheorem{assumption}{Assumption}
\newtheorem{definition1}{Definition}

\newcommand*{\tran}{^{\mkern-1.5mu\mathsf{T}}}
\newcommand*\dif{\mathop{}\!\mathrm{d}}

\DeclareMathOperator*{\argmin}{arg\,min}

\newcommand{\fakepar}[1]{\vspace{1mm}\noindent\textbf{#1.}}

\newcommand{\true}[1]{#1}

\newcommand{\sysc}{\theta}

\newcommand{\syscpr}{\theta}
\newcommand{\syscpo}{\theta}

\newcommand{\cov}{\Sigma}
\newcommand{\AB}{[A \; B]\tran}
\newcommand{\fpr}{\lambda}

\newcommand{\synth}{\text{\emph{synth}}}
\newcommand{\elength}{{\bar N^*}}

\newcommand{\switchk}{{\upsilon(k)}}

\newcommand{\credible}{\alpha}
\newcommand{\margin}{\xi}

\newcommand{\boundT}{\Delta_{\text{min}}}
\newcommand{\window}{\tau}
\newcommand{\fprcost}{\fpr \Delta N^*}

\newcommand{\ie}{i\/.\/e\/.,\/~}
\newcommand{\eg}{e\/.\/g\/.,\/~}
\newcommand{\cf}{cf\/.\/~}
\newcommand{\fig}{Fig\/.\/~}
\newcommand{\sect}{Sec\/.\/~}

\newcommand{\assum}{Assumption~}

\newtheorem{theorem}{Theorem}
\newtheorem{lemma}{Lemma}

\makeatletter
\newcommand\fs@betterruled{%
  \def\@fs@cfont{\bfseries}\let\@fs@capt\floatc@ruled
  \def\@fs@pre{\vspace*{5pt}\hrule height.8pt depth0pt \kern2pt}%
  \def\@fs@post{\kern2pt\hrule\relax}%
  \def\@fs@mid{\kern2pt\hrule\kern2pt}%
  \let\@fs@iftopcapt\iftrue}
\floatstyle{betterruled}
\restylefloat{algorithm}
\makeatother

\title{\LARGE \bf
Improving the Performance of Robust Control through \\ Event-Triggered Learning
}

\author{Alexander von Rohr\textsuperscript{1,2,3}, Friedrich Solowjow\textsuperscript{1,2}, and Sebastian Trimpe\textsuperscript{1,2}
\thanks{\raggedright\textsuperscript{1}Institute for Data Science in {Mechanical Engineering}, RWTH Aachen University,
        Aachen, Germany, vonrohr@dsme.rwth-aachen.de}
\thanks{\noindent\textsuperscript{2}Max Planck Institute for Intelligent Systems, Stuttgart, Germany}
\thanks{\noindent\textsuperscript{3}IAV GmbH Ingenieurgesellschaft Auto und Verkehr, Germany}
\thanks{\noindent
This work was supported in part by the Cyber Valley Initiative and the Max Planck Society.
The authors thank the International Max Planck Research School for Intelligent Systems for supporting A.~von~Rohr and F.~Solowjow.}
}

\begin{document}

\maketitle
\thispagestyle{fancy}
\pagestyle{empty}

\begin{abstract}
Robust controllers ensure stability in feedback loops designed under uncertainty but at the cost of performance. Model uncertainty in time-invariant systems can be reduced by recently proposed learning-based methods, which improve the performance of robust controllers using data. However, in practice, many systems also exhibit uncertainty in the form of changes over time, \eg due to weight shifts or wear and tear, leading to decreased performance or instability of the learning-based controller. 
We propose an event-triggered learning algorithm that decides \emph{when} to learn in the face of uncertainty in the LQR problem with rare or slow changes. Our key idea is to switch between robust and learned controllers.
For learning, we first approximate the optimal length of the learning phase via Monte-Carlo estimations using a probabilistic model. We then design a statistical test for uncertain systems based on the moment-generating function of the LQR cost. The test detects changes in the system under control and triggers re-learning when control performance deteriorates due to system changes. We demonstrate improved performance over a robust controller baseline in a numerical example.%
\end{abstract}

\section{Introduction}\label{sec:intro}

A common cause for poor control performance in industrial applications is long-term changes of the plant. 
After initial design and tuning, the controller is left unchanged for years \cite{jelali2006overview}.
When the dynamics of the plant change over time, performance degrades, and ad-hoc or nominally designed controllers might even become unstable.
In contrast, robustly designed controllers guarantee stability and worst-case performance for a pre-defined range of operating conditions \cite{zhou1996robust}.
This robustness against model uncertainties is not for free: it comes at the cost of performance \cite{boulet2007fundamental}.
Over a long enough period with slow or rare changes, a less robust controller that is routinely maintained and re-tuned will lead to a higher performance over the plant's life cycle.
Herein, we discuss how to automate controller re-tuning to improve performance while retaining robustness to model uncertainty.
More generally, finding time-dependent controllers can also be phrased as a time-varying convex optimization problem \cite{simonetto2020time} or a sequential decision problem \cite{brunzema2022controller}.

\begin{figure}[t]
\centering
\input{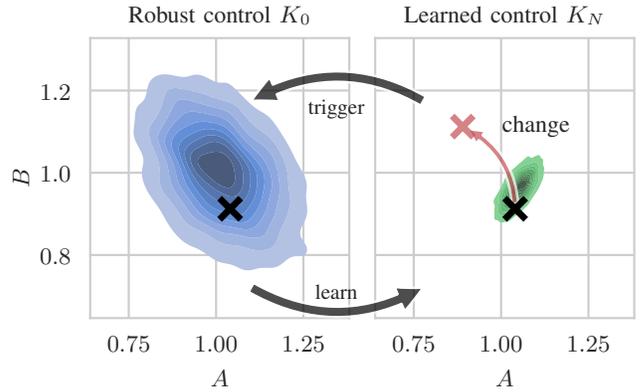}
\vspace{-5.5mm}
\caption{
\textbf{Overview of the event-triggered learning framework.} Initially, the current parameters of the system ($A$,$B$) have high uncertainty (left). Learning improves control performance by reducing uncertainty (right). Whenever the system leaves the learned uncertainty set, we detect the change~(red), fall back to the robust controller $K_0$ and learn again. 
}
\label{fig:fig1}
\end{figure}

We look at the re-tuning problem from the viewpoint of the recently introduced framework of event-triggered learning~(ETL)~\cite{solowjow2020event, schluter2020event, umlauft2019feedback}.
Event-triggered learning answers the question of \emph{when to learn} by using statistical tests to compare data-streams and models. 
Through learning, it is possible to update models and improve controllers.
However, successful learning requires excitation, which creates additional cost.
Therefore, learning should be limited to instances where the system's behavior deviates significantly from a learned model~(\cf\fig\ref{fig:fig1}).
For the question of how to learn, we build upon recent advances in learning-based robust control, which uses learned model uncertainty to design robust controllers (\cf \cite{berkenkamp2015safe, umlauft2018scenario, umenberger2018learning, rohr2021probabilistic} and~\citet{brunke2022safe} for a recent overview).
Essentially, reduced uncertainty through learning improves control performance.
We extend this idea by reducing uncertainty online in an event-triggered fashion.

A core assumption in most works on learning-based control is time-invariant dynamics.
In this static setting, an informative data-set collected \emph{offline} is sufficient to reduce uncertainty and improve control performance \emph{online}.
During offline learning, safety considerations, such as stability, might be relaxed, which is impossible in online learning.
However, when the system changes over time additional data collection must occur during online operations.
Then, the excitation signal for data collection incurs costs online and this cost needs to be amortized by the improved controller before the collected data becomes stale.
Stale data and excitation costs are not a problem in offline learning.

Herein, we consider the probabilistically robust LQR task with uncertain dynamics~\cite{berkenkamp2015safe,umenberger2018learning,rohr2021probabilistic} for systems with gradual or rare changes.
In this setting, we do not need robustness against all possible changes all the time.
If changes can be detected reliably, we can temporarily reduce model uncertainty through data and improve performance.
When a change is detected, we react by increasing the robustness.

Our proposed algorithm lies between time-invariant robust control and online adaptive control.
Adaptive control \cite{landau2011adaptive} updates parameters and controllers continuously online enabling the system to adapt to rapid changes. 
In environments with slow or rare changes, adaptive control can be problematic; for example, when data is not informative enough due to a lack of excitation, estimates can diverge~\cite{anderson2005failureadaptive}.
We consider rare changes and divide the plant life-cycle into episodes of unknown length with time-invariant dynamics characterizing each episode.
The proposed control scheme automatically detects changes via statistical tests.
In particular, we derive model-based confidence intervals to decide if online data-streams match the underlying model.
Our problem formulation relates to time-dependent switched systems\,\cite{liberzon2003switching}.
While control for switched systems tackles similar problems, it usually switches between a-priori known controller candidates.
In contrast, our focus is on learning-based control with probabilistic uncertainties.

\fakepar{Contributions}
The problem of detecting changes to a linear plant by designing a statistical test based on the~LQR cost has first been proposed by~\cite{schluter2020event}.
We extend this earlier work by generalizing the statistical test for the~LQR problem to incorporate model uncertainty.
Furthermore, we propose an~ETL-based robust control algorithm that uses this test to switch between a given conservative robust controller and online learned high-performance controllers.
We show that switching to learned controllers improves performance despite a-priori unknown changes in the environment.

\section{Preliminaries}

Consider a switched, linear, stochastic and discrete-time dynamical system
\begin{equation}\label{eq:linear_dynamics}
    x_{k+1} = \true A_\switchk x_{k} + \true B_\switchk u_{k} + \omega_k,
\end{equation}
where $x_k \in \mathbb{R}^{d_x}$ is the state, $u_k \in \mathbb{R}^{d_u}$ is the input, $\omega_k \sim \mathcal{N}(0,\true\cov_\switchk)$ is the process noise of the system and $\switchk \in \mathbb{N}$ is the switching signal.
We assume access to the state $x_k$ and i.i.d. process noise.
We define the system's unknown parameters as the tuple~$\theta_\switchk \dot= (A_\switchk, B_\switchk, \cov_\switchk)$.
\begin{assumption}\label{ass:episode}
We assume the system behaves episodically, that is, the switching signal changes only after an unknown dwell time $\Delta_i$.
The beginning of episode $i \in \mathbb{N}$ is denoted by $l_i$ and the dwell time is therefore given by $\Delta_i = l_{i+1} - l_{i}$.
Additionally, we define the counter $\switchk$ that tracks the number of switches
\begin{equation}\label{eq:switching_signal}
    \switchk = i, \quad \text{for } k \in [l_i,l_{i+1}).
\end{equation}
\end{assumption}

Note that we often omit the subscript $\switchk$ to simplify notation.
Here, we assume a normal distribution on the system's parameters, enabling closed-form Bayesian updates.
\begin{assumption}\label{ass:prior}
The system parameters $\true\theta_\switchk$ are sampled at the start of each episode $l_i$ with $l_0=0$ from a known (conjugate) prior distribution
\begin{equation}\label{eq:prior}
    \cov \sim \mathcal{W}^{-1}(V_0, v_0), \;\; \AB \sim \mathcal{MN}([\bar A \; \bar B]\tran, \Lambda^{-1}_0, \cov),
\end{equation}
where $\mathcal{W}^{-1}$ is the inverse Wishart distribution with parameters $V_0$ and $v_0$ and $\mathcal{MN}$ is the matrix normal distribution with mean $[\bar A \; \bar B]\tran$, row variance $\Lambda^{-1}_0$ and column variance\,$\cov$.
\end{assumption}

The probability density function of the system is denoted $p_{\theta}$.
We assume that, with high probability, a single state-feedback controller $K_0$ can stabilize samples from the prior.
\begin{definition1}[$\credible$-probabilistic robust controller]\label{def:apr}
A state feedback controller~$K$ is~$\credible$-probabilistic robust w.r.t.~$p_\theta$ if~$\mathbb{P}_{p_\theta}( \rho(A+BK) < 1) \geq 1 - \credible$, where $\rho(M)$ is the spectral radius of a matrix $M$.
\end{definition1}
\begin{assumption}\label{ass:stabilizable}

We assume the prior $p_\theta$ is $\credible$-probabilistic stabilizable, \ie there exists a state feedback controller $K_0$ that is $\credible$-probabilistic robust w.r.t. $p_\theta$.  
\end{assumption}

In practice, an $\credible$-probabilistic robust controller can be found by stabilizing all systems inside a set $\Theta_\credible$ defined by an $\credible$ credible region.
We synthesize $K$ using an algorithm, which we denote as 
\begin{equation}\label{eq:synth}
\text{synth} : \Theta_\credible \xrightarrow[]{} \mathcal{K} \subseteq \mathbb{R}^{d_u \times d_x}.
\end{equation}
Many such algorithms with different performance criteria and uncertainty considerations have been developed in the robust control community (\eg \cite{zhou1996robust,calafiore2006scenario}).
Here, we choose a worst-case design proposed by \citet{berkenkamp2015safe}.

To enable online learning, we assume the time between changes is long enough, leading to Assumption\,\ref{ass:T}.
\begin{assumption}\label{ass:T}
We assume the dwell time $\Delta_i$ is lower bounded by a known constant~$\boundT \in \mathbb{N}^+$, $\Delta_i \geq \boundT$ and~$\boundT \gg m$ where~$m$ is the frequency of sub-sampling that avoids correlation in the data (\cf \cite{shiranifaradonbeh2018finite,simchowitz2018learning}). 
\end{assumption}

This assumption allows us to collect sufficient data to improve the controller.
A minimum dwell time enables learning-based techniques to identify the system and improve performance.
At the same time, leveraging statistical tests and updating the state-feedback only when necessary avoids constant excitation.
Further, \assum\ref{ass:T} allows us to ignore effects associated with switched dynamics. since changes and controller updates only happen sporadically.
It follows directly from \assum\ref{ass:stabilizable} that any $K_0$ will $\credible$-stabilize the time-varying dynamics of \eqref{eq:linear_dynamics} \cite[Thm.\,3.2]{liberzon2003switching}.

To exploit the knowledge about $\boundT$, we leverage data to update a probabilistic dynamics model presented in the following subsection.

\subsection{Bayesian Model Update with Trajectory Data}\label{subsec:update}

The Bayesian dynamics model update of $\theta$ can be stated as standard Bayesian multivariate linear regression, which, using the prior in \eqref{ass:prior}, yields a closed-form update.
The estimation problem for $\AB$ and $\cov$  is 
\begin{equation}
    Y = X \AB + E,
\end{equation}
with 
\begin{equation}
X = \begin{bmatrix}  
x\tran_0 & u\tran_0\\  
x\tran_{m-1} & u\tran_{m-1}\\  
\vdots & \vdots \\
x\tran_{N m -1} & u\tran_{N m -1}\\  
\end{bmatrix}
\;
Y = \begin{bmatrix}  
x\tran_1 \\ 
x\tran_{m} \\
\vdots \\
x\tran_{N m}\\  
\end{bmatrix}
\;
E = \begin{bmatrix}  
\omega\tran_0 \\
\omega\tran_{m-1} \\
\vdots \\
\omega\tran_{N m -1}\\  
\end{bmatrix},
\end{equation}
where $X$ and $Y$ contain the $m$-sub-sampled trajectory data, and $E$ is the unobserved error matrix containing the disturbances.
We perform sub-sampling to obtain (approximately) i.i.d. data, as required by the estimator and justified in \cite[Lemma 5]{schluter2020event}.
A sufficient sub-sampling interval, called mixing time, can be estimated from data \cite{solowjow2020kernel}.

The closed-form update of Bayesian linear regression requires the inverse $(X\tran X)^{-1}$ to exist and, thus, $X$ to have full rank.
However, for data from the closed-loop system, $x_k$ and $u_k$ are linearly dependent.
This problem is well known in system identification and is addressed by an additional input signal~$e$ exciting the system. The resulting system is
\begin{equation}\label{eq:excitation}
    x_{k+1} = (\true A + \true B K) x_{k} + \true B  e_{k} + \omega_k,
\end{equation}
where $e_{k}$ is sampled i.i.d. from a stochastic excitation policy $e_k \sim \pi_e$, which we pick normal.
The control cost will then remain finite if $K$ is stabilizing, as we prove in\,{\sect\ref{sec:improve}}.

\subsection{Problem Setting}
Given \eqref{eq:linear_dynamics} and a robust synthesis algorithm \eqref{eq:synth}, we can use data to update the Bayesian distribution.
We now design an algorithm to minimize the expected LQR cost by reducing uncertainty over time.
We restrict the formulation to one update of the state feedback~$K$ per episode. 
The expectation of the quadratic cost over a time window of size~$\window \in \mathbb{N}^+$ with respect to a probabilistic model~$p_\theta$ is
\begin{subequations}\label{eq:cost_all}
\begin{align}
\mathbb{E}_\theta [J_k(K)] & = \int_{\Theta_\credible} J_k(K, \theta) p_\theta(\theta) \dif \theta, \label{eq:cost_int} \\
J_k(K,\theta)& = \sum^\window_{j=k} \mathbb{E}_{\omega}\left[ x_j^T Q x_j  + u_j^T R u_j\right] \label{eq:cost_state} \\
\text{s.t.} \quad x_{j+1} &= (A + BK) x_{j} + \omega_j, \; \theta = [A \; B \; \cov] \nonumber
\end{align}
\end{subequations}
where $Q$ and $R$ are user-defined, positive definite weight matrices, and the initial condition is sampled from the stationary distribution of the closed loop system~$x_k \sim \mathcal{X}$.
We consider two types of uncertainty for the cost: the regular LQR stochasticity w.r.t. the process noise, and parameter uncertainty.
More specifically, any fixed value for $\theta$ yields the LQR setting, but the parametric uncertainty in $\theta$ makes \eqref{eq:cost_state} a random variable. 
To obtain a deterministic object for the cost, we consider the expectation of that random variable over a bounded domain.
For \eqref{eq:cost_int} to exist, $K$ must stabilize almost all systems in $\Theta_\credible$. 

Our goal is to improve the performance of a robust controller $K_0$ through learning.
For this we propose to learn only when necessary, and for a limited time.
Determining the learning phase's length is critical because excitation is costly and uncertainty reduction is only temporary.

We design an~ETL algorithm with a learning phase of variable length $N$ per episode $\Delta_i$ and total expected cost
\begin{align}\label{eq:total_cost}
    &\mathbb{E}[J_{\Delta_i} (N)] = N \mathbb{E}_{\syscpr}[J_{\pi_e}(K_0)] + (\Delta_i-N) \mathbb{E}_{\syscpo}[J(K_N)] 
\end{align}
where $K_N$ is the improved $\credible$-probabilistic robust controller after $N$ excitation steps.
The first term $\mathbb{E}_\syscpr [J_{\pi_e}(K_0)]$ is the expected cost of system \eqref{eq:excitation} using the excitation signal $\pi_e$ w.r.t. $p_\syscpr$ under the robust state-feedback $K_0$.
The additional cost of excitation is $\Psi = \mathbb{E}_\syscpr[J_{\pi_e}(K_0)] - \mathbb{E}_\syscpr[J(K_0)]$ which needs to be amortized by the improvement after learning.

The expected cost after the improvement $\mathbb{E}_{\syscpo}[J(K_N)])$ is
\begin{equation}
    \mathbb{E}_{\syscpo}[J(K_N)] = \mathbb{E}_{\syscpo}[\mathbb{E}_{X}[J(K_N) \mid \theta ]].
\end{equation}
Given the system parameters $\theta$, the feedback matrix $K_N$ is a random variable determined by the collected data $X$.
From a practical point of view, this decomposition of expectation can be used to approximate the improved cost via sampling.

The optimal cost is given by minimizing over $N$ 
\begin{align}\label{eq:opt_length_delta}
    N^* &= \argmin_{N \in (0,\Delta_i)} \mathbb{E}[J_{\Delta_i}(N)].
\end{align}
Since $\Delta_i$ is unknown, we must detect the end of an episode characterized by a change in the dynamics.
To this end, we design a trigger that activates the controller $K_0$ and learning.

\section{Improving a Probabilistic Robust Controller}

In this section, we improve the expected performance of a baseline $\credible$-probabilistic robust controller by reducing uncertainty through learning.
First, we design the learning experiment by estimating the optimal excitation length.
Second, we detect changes in the dynamics using statistical tests.
Third, we show the achieved performance improvement. 

\subsection{How Long to Learn?}\label{sec:optimal_e}

To determine the optimal length of the excitation $N^*$, we need to solve \eqref{eq:opt_length_delta} which is dependent on the episode length.
Since~$\Delta_i$ is unknown a-priori, we utilize the lower bound~$\boundT$ and we obtain an upper bound on the cost by optimizing
\begin{align}\label{eq:opt_length}
    \elength &= \argmin_{N \in (0,\boundT)} \mathbb{E}[J_{\boundT}(N)]. 
\end{align}
If $\elength = 0$, the current controller is already performing sufficiently. Exciting the system would only yield additional costs which the improved controller cannot amortize.
Thus, there is no learning, and we recover the initial robust controller.
Next, we show that we minimize an upper bound for the cost by solving \eqref{eq:opt_length}. 
\begin{lemma}\label{lemma:lower_bound}
Let Assumption~\ref{ass:T} hold then
\begin{equation}
    \nicefrac{1}{\boundT} \; \mathbb{E}[J_{\boundT}(N)] \geq \nicefrac{1}{\Delta} \; \mathbb{E}[J_{\Delta}(N)].
\end{equation}
The expected cost of over an episode is lower bounded by the cost of the shortest episode.
\end{lemma}
\begin{proof}
The proof follows directly from \assum\ref{ass:T}.
\begin{equation*}
\begin{aligned}
    \nicefrac{1}{\boundT} \; N (\mathbb{E}[J_{\pi_e}(K_0)] &- \mathbb{E}[J(K_N)]) + \boundT  \mathbb{E}[J(K_N)] 
    \\
    \geq \nicefrac{1}{\Delta} \; N (\mathbb{E}[J_{\pi_e}(K_0)] &- \mathbb{E}[J(K_N)]) + \Delta  \mathbb{E}[J(K_N)] \\ 
    \Rightarrow \boundT &\leq \Delta 
\end{aligned}
\vspace{-4mm}
\end{equation*}
\end{proof}
Next we describe how to estimate the needed expectations in \eqref{eq:total_cost} via MC integration.

\fakepar{Cost of excitation $\mathbb{E}_\syscpr[J_{\pi_e}(K_0)]$}
The expected excitation cost $\mathbb{E}[J_{\pi_e}(K_0)]$ consists of two nested expectations in \eqref{eq:cost_all}.
For closed-loop linear systems \eqref{eq:linear_dynamics} with quadratic cost, the inner expectation \eqref{eq:cost_state} (w.r.t. the noise) can be computed efficiently by solving a convex optimization problem with~LMI constraints \cite[Lemma 2]{schluter2020event}.
Next, we show how to determine the cost for the fixed excited system \eqref{eq:excitation}.
\begin{theorem}\label{th:excitation_cost}
Given positive definite $Q$ and $R$, the expected value for the quadratic cost of the excited system \eqref{eq:excitation} over a window of size $\tau$ with excitation signal $e_k \sim \mathcal{N}(0,\Sigma_e)$ is
\begin{equation}\label{eq:ex_cost}
    \mathbb{E}_{\omega}[ J_{\pi_e}(K, \theta) ]  = J_k(K, \tilde \theta) + \window \; \textup{tr} \left({\Sigma_e R }\right),
\end{equation}
where $\tilde \theta = ( A,B,\cov + B \Sigma_e B^T )$ and $J_k(K, \tilde \theta)$ as in \eqref{eq:cost_state}. 
\end{theorem}
\begin{proof}
First, we determine the expectation of a squared white-noise signal weighted with $R$ (and $\window=1$),
\begin{equation}
    \mathbb{E} [e_k\tran R e_k] = \mathbb{E} [ \textup{tr} (e_k\tran e_k R)] = \textup{tr} (\cov_e R),
\end{equation}
which is the additional input cost caused by the excitation.

The input signal induces an additional cost term by influencing the system's state.
Because the linear transformation and sums of multivariate normal random variables are still normal, the system can be rewritten as
\begin{align}
    x_{k+1} &= (\true A + \true B K) x_{k} + \tilde e_{k} + \omega_k, \\
    x_{k+1} &= (\true A + \true B K) x_{k} + \tilde \omega_k,
\end{align}
with $\tilde e_{k} \sim \mathcal{N}(0, B \Sigma_e B^T )$ and $\tilde \omega_{k} \sim \mathcal{N}(0, \cov + B \Sigma_e B^T )$.
The claim then follows from \cite[Lemma 2]{schluter2020event}.
\end{proof}
We approximate the expectation w.r.t. to $\theta$, $\mathbb{E}_{\theta}[ J_{\pi}(K_0) ]$, via MC integration. 
We are still missing the estimation of the improved cost $\mathbb{E}[J(K_N)]$ which, in contrast to $\mathbb{E}_{\omega}[ J_{\pi_s}(K, \theta) ] $, is dependent on the variable excitation length $N$.

\fakepar{Estimating the learning outcome $\mathbb{E}_{\syscpo}[J(K_N)]$}\label{subsubsec:learning}
Predicting the improvement when reducing the uncertainty of a Bayesian model through data is analytically intractable.
Therefore, we resort to  MC integration to estimate the expectations required to improve robust control performance.

\begin{figure}[t]
\centering
\vspace{1mm}
\input{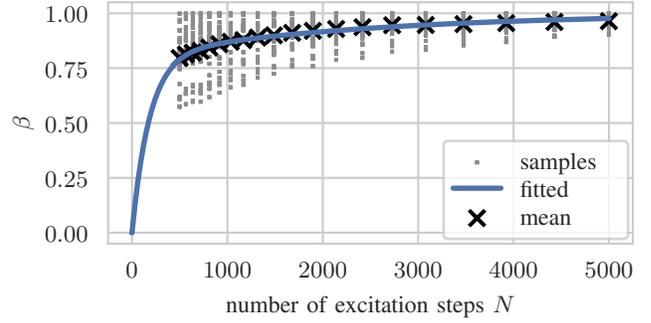}
\vspace{-5.5mm}
\caption{\textbf{Improvement rate $\beta(N)$ for a 1D-Example prior:} The parameterized $\beta(N)$ is fitted to the empirical mean of 25 sampled improvement rates. Each point represents a cost of an $\credible$-probabilistic robust controller using $N$ data points for the Bayesian posterior. Due to a high variance for low $N$, we start the MC simulation at $N = 250$ letting the parameterized model interpolate between $N=0$ and $N=250$ (\cf \sect\ref{subsec:emperical_length}).}
\label{fig:est_beta}
\end{figure}

Our goal is to estimate the rate of cost improvement as a function of $N$, which is illustrated in \fig\ref{fig:est_beta}.
We do this by fitting an ad-hoc parameterized function $\beta(N)$ to the MC samples.
Here, the function should satisfy $\beta(0) = 0$ and $\lim_{N \rightarrow \infty} \beta(N) = 1$.
We choose heuristically 
\begin{equation}
    \beta(N) = \gamma_1 (1 - \gamma_2^{-\gamma_3 N}) + (1-\gamma_1) (1 - \gamma_4^{-\gamma_5 N}),
\end{equation}
with~$\gamma_1 \in (0,1)$ and~$\gamma_{2},\dots,\gamma_{5} > 0$.
The improvement rate is defined as
\begin{equation}
    \mathbb{E}[J(K_N)] = \mathbb{E}_{\syscpr}[J(K_0)] - \beta(N) \mathbb{E}_{\syscpr}[G(K_0)],
\end{equation}
where $G(K_0)$ is the sub-optimality gap of a robust controller
\begin{equation}\label{eq:gap}
    G(K, \theta) = J(K, \theta) - J(K_\theta, \theta).
\end{equation}
We can compute $J(K, \theta)$ as in \cite[Lemma 2]{schluter2020event} and determine~$K_\theta$ by solving the corresponding Riccati equations.
We estimate $\beta(N)$ by
\begin{enumerate*}
    \item sampling systems $\theta$ from the prior distribution \eqref{eq:prior};
    \item generating a data set $X$ of size $N$;
    \item calculating the posterior;
    \item synthesizing an $\credible$-robust controller $K_N$; and
    \item analytically determining $G(K_N, \theta)$ using the sampled system parameters.
\end{enumerate*}
We repeat this for several samples $\sysc$, lengths $N$, and data sets~$X$, thereby sampling from $\beta(N)$ using only the known \emph{prior} distribution.
Because the improvement rate is a property of the prior and episode length, the excitation signal can be calculated offline.
Next, we derive a statistical test that detects changes to guarantee closed-loop stability and cost improvement.

\subsection{When to Learn?}\label{sec:detect}

After the improvement, the controller is $\credible$-probabilistic robust w.r.t. the new posterior.
However, this comes at the price of reduced robustness against future changes.
While the prior controller $K_0$ can stabilize all systems in the prior uncertainty set, the posterior controller $K_\elength$ improves performance using the smaller posterior set given the generated data.
In the event of a change, the controller's performance can degrade or, in the worst case, become unstable.

An online algorithm must detect such changes reliably while tolerating noisy observations.
When change is detected, the algorithm can return to the robust controller and restart learning.
The overall goal of our framework is guaranteed improvement in expectation. We design a statistical test that detects significant deviations in the cost and, thus, the potential for improvement. 
Since learning is costly, this that controls the number of false positives due to noise, \ie triggering when no change is present.

The test is based on prior work \cite{schluter2020event}, which derives confidence bounds on the LQR cost for a fixed system.
Here we consider a distribution over systems via the Bayesian posterior.
The original test uses sharp Chernoff bounds to determine high probability confidence intervals $(\kappa^-, \kappa^+)$ for the measured and stochastic cost $\hat J_k$ given the system parameters $\theta$ and thus,
\begin{equation}\label{eq:bound}
    \mathbb{P}_{\omega}[ \hat J_k \not\in (\kappa^-, \kappa^+) \mid \theta] \leq \eta,
\end{equation}
where $k$ is the start time of the current window, $\hat J_k$ is the observed cost over that window, and $\eta$ is a (low) risk parameter (\cf\cite[Theorem 3]{schluter2020event}).
The probability is given w.r.t. the noise.
To generalize the test, we define an indicator variable 
\begin{equation}
\phi(k) = \begin{cases}
                1, & \text{if} \;\; \hat J_k \not\in (\kappa^-, \kappa^+)  \\
                0, & \text{otherwise.}
            \end{cases}.
\end{equation}
In \citet{schluter2020event}, the system parameters are point estimates, whereas we have a distribution over system parameters. 
Therefore, we also obtain a distribution over~$(\kappa^-, \kappa^+)$ and the indicator variable $\phi(k)$.
We replace the indicator with its expectation, since  $\mathbb{P}_{\omega}[\phi(k) = 1] = \mathbb{E}_{\hat J}[\phi(k)]$, and define a (high) confidence level $\nu$ so that
\begin{equation}\label{eq:confidence}
\mathbb{P}_\theta[\mathbb{E}_{\omega}[\phi(k) \mid \theta] > \eta] \leq 1 - \nu.  
\end{equation}
In practice, we first estimate a distribution over the bounds by sampling systems from the \emph{posterior} and pick the appropriate $\nu$-percentile to obtain $(\kappa_\nu^-, \kappa_\nu^+)$.

Finally, the trigger to switch to the fallback controller and start a new learning experiment is given by
\begin{equation}\label{eq:trigger}
    \mathbb{P}[ \hat J_k \not\in (\kappa_\nu^-, \kappa_\nu^+) ] \leq 1 - \nu (1-\eta),
\end{equation}
which can be evaluated using the observed cost and ensures~\eqref{eq:confidence}.
If the observed cost leaves $(\kappa_\nu^-, \kappa_\nu^+)$, Algorithm~\ref{alg:betl} switches to $K_0$ and triggers learning.
The false positive rate~$\fpr$ of the trigger follows directly from \eqref{eq:trigger} as $\fpr \leq 1 - \nu (1-\eta)$.

\begin{figure*}[t]
\vspace{2mm}
\centering
\input{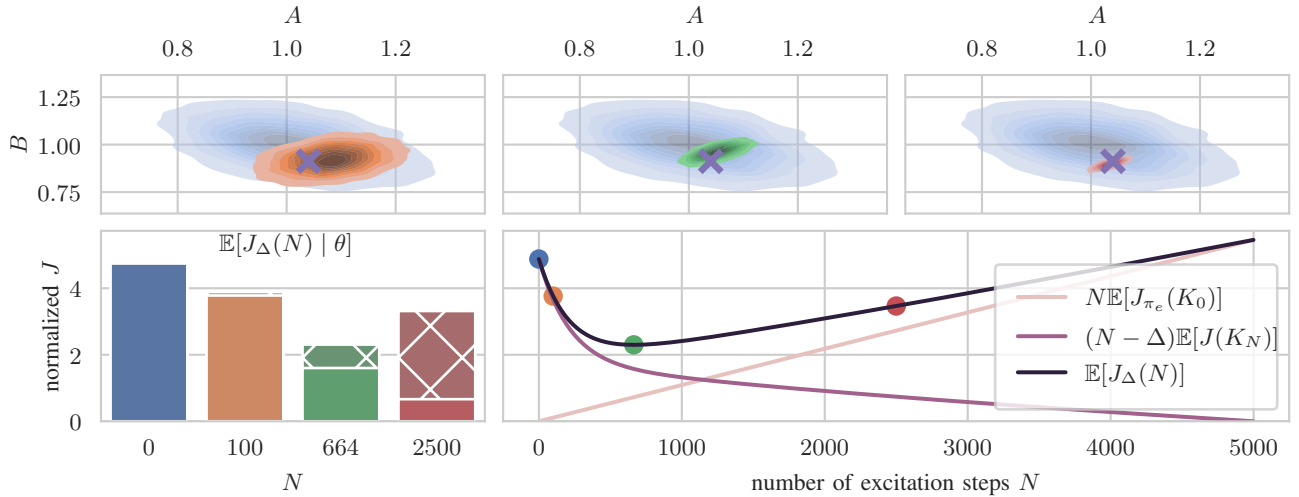}
\vspace{-5.5mm}
\caption{\textbf{Different excitation lengths for a 1D-Example prior:} In the top row, we show the prior distribution (light blue) and the posterior distribution over~$\AB$ for different excitation signal lengths $N$.
On the lower left, we see the total cost over one episode \eqref{eq:total_cost} including regular operation and excitation phase (hatched).
The lower right shows the estimated expectation of the cost using the prior distribution, but no data from the ground truth system (purple cross, top row).
The optimal expected cost (green) yields improved total cost on the ground truth.
After $N=100$ data points (yellow), the posterior variance is still high, and the resulting controller has a sizeable sub-optimality gap.
In contrast, after $N=2500$ the variance has almost vanished, and the controller is close to optimal, but there is not enough time left in the episode to amortize the excitation cost.}
\label{fig:example}
\end{figure*}

\subsection{Improvement}\label{sec:improve}

\begin{algorithm}[t]
    \caption{Improving robust control}\label{alg:betl}
	\begin{algorithmic}[1]
		\State \textbf{Input:} Parameters of the prior \eqref{eq:prior}, $\credible$-probabilistic robust controller $K_0$, synthesis algorithm $\synth$, trigger parameter $\eta$, $\nu$ and $\margin$, and minimum dwell time $\boundT$.
		\State Estimate improvement rate $\beta(N)$ (\sect\ref{subsubsec:learning})
		\State Approximate the optimal signal length $\hat N^*$ (\sect\ref{sec:optimal_e})
        \For{each episode~$i$}
            \State $K \leftarrow K_0$ \Comment{Use robust controller.} 
    		\State Apply the excitation signal for $\hat N^*$ steps.
    		\State Update model and credible region $\Theta_\credible$. (\sect\ref{subsec:update})
            \State Estimate trigger bounds $(\kappa_\nu^-, \kappa_\nu^+)$. (\sect\ref{sec:detect})
                \State $K \leftarrow \synth(\Theta_\credible)$ \Comment{Use learned controller.}
            \Do
                \State Observe cost $\hat J_w(k)$ of learned controller. 
            \doWhile{$\hat J_w(k) \in (\kappa_\nu^- - \margin, \kappa_\nu^+ + \margin)$ }
        \State Reset model to prior. \Comment{Episode end detected.}
        \EndFor
	\end{algorithmic}
\end{algorithm}

Next, we show that the proposed algorithm improves the expected cost.
The additional cost decomposes into the learning cost $\Psi$, late-detection cost, and false positives cost of the trigger.
The expected number of false positives in an episode of length $\Delta$ is at most $\fpr \Delta$ and the cost of each trigger is $\elength \Psi$.
Therefore, the additional expected cost for an episode at most $\fpr \Delta \elength \Psi$.
We deal with false negatives by ensuring that the trigger thresholds are sufficiently below the cost of the robust controller. 

Furthermore, we require an additional assumption that quantifies how fast the cost increases after a system change.
\begin{assumption}\label{ass:trigger_cost}
For all $K \in \mathcal{K}$ (see \eqref{eq:synth}), if $J_k(K) \leq \kappa_\nu^+$
\begin{equation}
    \mathbb{E} [J_{k+\window}(K)]  \leq \Omega.
\end{equation}
\end{assumption}

If at time $k$, the cost is below the trigger threshold, we assume that it remains bounded until the next test, even if the controller is unstable. 
This assumption enables amortizing the cost incurred before detecting a system change.
In principle, one could trigger the fallback controller before the cost window is over to enforce this assumption, \eg using thresholds on the system's states. 
In our empirical evaluations, this cost is always small.

We are now ready to state the final result---expected cost improvement with high probability:
\begin{theorem}\label{th:stability}
Let Assumption~\ref{ass:episode}-\ref{ass:trigger_cost} hold, $\elength$ be the solution to \eqref{eq:opt_length}, and
$\mathbb{E}[\kappa_\nu^+]$ be the expected value of the upper trigger threshold.
Set a margin on the expected trigger threshold
$$
\margin = \frac{(1+\fpr \boundT) \elength + \Omega}{\boundT - (1 + \fpr \boundT) N^* } \Psi,
$$
where $\Psi = \mathbb{E}[J_{\pi_e}(K_0) - J(K_0)]$ is the additional cost of one learning experiment over the baseline controller $K_0$.

Algorithm~\ref{alg:betl} ensures with probability $\credible$ that
\begin{equation}\label{eq:result}
     \mathbb{E}[J_{\Delta}(\elength)] \leq  \Delta \mathbb{E}[J(K_0)],
\end{equation}
if $\mathbb{E}[\kappa_\nu^+] + \margin \leq \mathbb{E}[J(K_0)]$.
\end{theorem}
\begin{proof}
If $\elength=0$ then $\mathbb{E}[J_{\Delta}(\elength)] =  \Delta \mathbb{E}[J(K_0)]$.

Under the assumption $\boundT \leq \Delta$, the margin gets smaller with a longer episode,
\begin{equation}
\margin \geq \frac{(1+\fpr \Delta) \elength + \Omega}{\Delta - (1 + \fpr \Delta) \elength} \Psi.
\end{equation}
The total cost of Algorithm~\ref{alg:betl} given the episode  length $\Delta$ is   
\begin{align*}
    \mathbb{E}[J_{\Delta}(\elength)] \leq 
    &\underbrace{\elength (\Psi + \mathbb{E}[J(K_0)])}_{\text{true positive}} 
    + \underbrace{\fprcost (\Psi + \mathbb{E}[J(K_0)])}_{\text{false positive}} \\
    &+ \underbrace{(\Delta - \elength - \fprcost) \mathbb{E}[\kappa_\nu^+]}_{\text{false and true negative}} \\
    &+ \underbrace{\mathbb{E}[J_k(K_{\elength}) \mid J_k(K_{\elength}) > \kappa_\nu^+]}_{\text{trigger delay}}\\
    \leq & (1+\fpr \Delta) \elength  (\Psi + \mathbb{E}[J(K_0)] \\
    &+ (\Delta - (1+\fpr \Delta) \elength) (\mathbb{E}[J(K_0)] - \margin) + \Omega \\
    \leq &\Delta \mathbb{E}[J(K_0)] \\
    &- (1+\fpr)\elength \Psi + (1+\fpr) \elength \Psi - \Omega + \Omega \\
    \leq &\Delta \mathbb{E}[J(K_0)].
\end{align*}
The probability $\credible$ follows from the fact that the expectations are conditioned on the event that $\theta \in \Theta_\credible$.
\end{proof}
%
In practice, we estimated the expectation using MC sampling from the prior.
In the next section, we demonstrate that these estimates are sufficient for the algorithm to achieve substantially improved cost over the robust baseline.

\section{Empirical examples}

This section contains two illustrating simulation examples.\footnote{Python implementation at \url{https://github.com/avrohr/betl}.}
By means of a one-dimensional example, we first highlight how the uncertainty about the system parameters relates to the resulting cost of an $\credible$-probabilistic robust controller. We estimate the optimal length of an excitation signal to optimize performance over an episode.
In the second part, we illustrate the complete algorithm on an instance of the LQR problem with unknown dynamics proposed by \citet{dean2020sample} and show a substantial cost improvement.

\subsection{Optimal Excitation Signal Length}\label{subsec:emperical_length}

This subsection illustrates the influence of the length of the excitation signal $N$ on the total cost over a fixed horizon.
For clarity, we choose a one-dimensional state and input space, which enables us to plot prior and posterior over the parameters $\AB$.
The prior distribution \eqref{eq:prior} is depicted in~\fig\ref{fig:example} and parameterized as
\begin{equation}\label{eq:1d_dynamics}
\begin{aligned}
\bar A &= 1.01, \;  \bar B = 0.1, \; \Sigma = 0.02, \; Q = 1, \; R= 100, \; \\
\Lambda^{-1}_0 &= 0.7 \, I_2 + 0.2 \, 1_2, \; v_0=100, \; V_0=0.02, \;
\end{aligned}
\end{equation}
where $I_i$ denotes the $i$-dimensional identity matrix and $1_i$ denotes a $i \times i$ matrix filled with ones. 
The credible region is set to $\credible=0.99$, and the ground truth $\theta$ was sampled from the prior.
The excitation variance is set to $\Sigma_e = 0.02$ and the episode length is given as $\boundT = 5000 \cdot m$.
We did not estimate the mixing time but set $m=20$ and verified the convergence of the estimator empirically.
To estimate the improvement rate, we sampled $25$ systems from the prior, determined the posterior control performance with different excitation signal lengths and fit $\gamma$ using a least-squares estimation.
The resulting improvement rate is depicted in~\fig\ref{fig:est_beta}.
Solving~\eqref{eq:opt_length} the optimal excitation signal length is estimated to be~$\elength=664$.
We compare the total cost over an episode with~$N=0$,~$100$ and~$2500$ where~$\elength$ yields the lowest total cost, as expected.
\fig\ref{fig:example} shows that $100$ samples of a trajectory still lead to high uncertainty and thus higher cost, while $2500$ samples lead to lower uncertainty and lower control cost after learning; however excitation and opportunity costs are high, leading to a higher total cost.

We want to emphasize that $\elength$ is estimated using only the prior: the estimator has no access to the ground truth nor to any samples from it.
\fig\ref{fig:example} shows the prior and posterior for the different $N$, the cost estimation, and the actual cost of the learned controller.
However, $\elength$ is optimal only in expectation, so for any actual realization of the system, the optimal signal can deviate.

\subsection{Improving Robust Control}

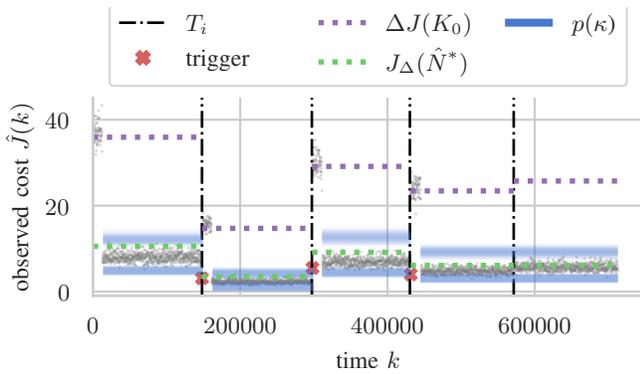
\begin{figure}[t]
\vspace{2mm}
\centering
\input{figures/improve-3d.pgf} 
\vspace{-5.5mm}
\caption{\textbf{Cost} (grey circles) \textbf{over time} of the proposed controller improvement scheme. System changes (vertical, black) are detected via the learning trigger \eqref{eq:trigger}. The statistical test quickly detects system changes (red) and triggers a learning phase. The last change is not detected since the cost of the changed system is inside the trigger bounds (blue). The algorithm avoids the costly fallback controller at the potential opportunity cost.
Compared to the robust controller (purple), the improved controller's total cost (green) is substantially lower.}
\label{fig:improve}
\end{figure}

As a benchmark problem, we use a prior distribution \eqref{eq:prior} for an LQR benchmark problem based on \citet{dean2020sample}
\begin{equation}\label{eq:dean_dynamics}
\begin{aligned}
A_0 &=
\begin{bmatrix}
    1.01 & 0.01 & 0\\
    0.01 & 1.01 & 0.01\\
    0 & 0.01 & 1.01
\end{bmatrix}, \;
\begin{aligned}
B_0 &= 0.1 \, I_3, & \Sigma = 10^{-3}\,I_3, \\
R &= \,I_3, & Q = 10^{-3}\,I_3,  \\
\end{aligned} \\
\Lambda^{-1}_0 &= 0.07 \, I_6 + 0.03 \, 1_6, \; v_0=10, \; V_0=6 \cdot 10^{-3} \,I_3. \;
\end{aligned}
\end{equation}
The lower bound of the episode length is $\boundT = 10000 \cdot m$ with mixing time $m=20$ and we randomize the episode lengths $\Delta_i$.
The other parameters are the same as in \sect\ref{subsec:emperical_length}.
The optimal excitation length for this prior is estimated to be $\elength = 549$. Since we sub-sample the system at a rate of $1/m$, the resulting learning phase in \fig\ref{fig:improve} is of length $\elength m$.
For the trigger, we set the number of MC samples to $50$, the cost window size $\window=200$, $\eta=0.002$ and $\nu=0.05$.

As seen in \fig\ref{fig:improve}, the controller cost can be improved after learning, and the true total cost of the controller is at least $3$ times lower than the cost of a static robust controller.
After a period of costly excitation, the learned controller substantially increases performance.
Further, we can reliably detect system changes if the cost of the new system differs from the old one.
In principle, the trigger can be made more sensitive to changes by adjusting $\eta$ and $\nu$, but this results in a higher false-positive rate.
Instead, one should optimize for a low margin $\margin$ (\cf Theorem~\ref{th:stability}).

\section{Conclusion}

In summary, we have shown how ETL can be used to improve probabilistic robust LQR performance for switched dynamics under probabilistic uncertainty.
The resulting algorithm makes two decisions specific to this problem setting: \emph{when to learn} and \emph{for how long}.
First, based on the current system model and performance of the learned controller, we decide whether the system has changed.
We extend a statistical test \cite{schluter2020event}
that directly operates on the LQR cost to take additional model uncertainty into account.
Second, we derive the optimal excitation length based on the prior model and a lower bound on the episode length.
We then optimize the length by estimating the improvement via MC simulations.
In empirical results, this substantially improves control performance of ETL over a robust baseline.

ETL bridges the fundamental trade-off between robustness and performance by reducing uncertainty through learning and statistical tests that detect changes.
For future work, we are interested in providing regret bounds, \ie quantifying the sub-optimality gap that ETL-based control algorithms can achieve when faced with probabilistic uncertainty.

\section*{Acknowledgment}
The authors thank D.~Baumann, P.~Brunzema, H.~Hose, J.~Schmacka, and P.-F.~Massiani for their helpful comments.

\printbibliography

\end{document}

%% file: figures/improve-3d.pgf
\begingroup%
\makeatletter%
\begin{pgfpicture}%
\pgfpathrectangle{\pgfpointorigin}{\pgfqpoint{3.390065in}{1.885658in}}%
\pgfusepath{use as bounding box, clip}%
\begin{pgfscope}%
\pgfsetbuttcap%
\pgfsetmiterjoin%
\definecolor{currentfill}{rgb}{1.000000,1.000000,1.000000}%
\pgfsetfillcolor{currentfill}%
\pgfsetlinewidth{0.000000pt}%
\definecolor{currentstroke}{rgb}{1.000000,1.000000,1.000000}%
\pgfsetstrokecolor{currentstroke}%
\pgfsetdash{}{0pt}%
\pgfpathmoveto{\pgfqpoint{0.000000in}{0.000000in}}%
\pgfpathlineto{\pgfqpoint{3.390065in}{0.000000in}}%
\pgfpathlineto{\pgfqpoint{3.390065in}{1.885658in}}%
\pgfpathlineto{\pgfqpoint{0.000000in}{1.885658in}}%
\pgfpathlineto{\pgfqpoint{0.000000in}{0.000000in}}%
\pgfpathclose%
\pgfusepath{fill}%
\end{pgfscope}%
\begin{pgfscope}%
\pgfsetbuttcap%
\pgfsetmiterjoin%
\definecolor{currentfill}{rgb}{1.000000,1.000000,1.000000}%
\pgfsetfillcolor{currentfill}%
\pgfsetlinewidth{0.000000pt}%
\definecolor{currentstroke}{rgb}{0.000000,0.000000,0.000000}%
\pgfsetstrokecolor{currentstroke}%
\pgfsetstrokeopacity{0.000000}%
\pgfsetdash{}{0pt}%
\pgfpathmoveto{\pgfqpoint{0.474609in}{0.377132in}}%
\pgfpathlineto{\pgfqpoint{3.356164in}{0.377132in}}%
\pgfpathlineto{\pgfqpoint{3.356164in}{1.414243in}}%
\pgfpathlineto{\pgfqpoint{0.474609in}{1.414243in}}%
\pgfpathlineto{\pgfqpoint{0.474609in}{0.377132in}}%
\pgfpathclose%
\pgfusepath{fill}%
\end{pgfscope}%
\begin{pgfscope}%
\pgfpathrectangle{\pgfqpoint{0.474609in}{0.377132in}}{\pgfqpoint{2.881555in}{1.037112in}}%
\pgfusepath{clip}%
\pgfsetroundcap%
\pgfsetroundjoin%
\pgfsetlinewidth{0.803000pt}%
\definecolor{currentstroke}{rgb}{0.800000,0.800000,0.800000}%
\pgfsetstrokecolor{currentstroke}%
\pgfsetdash{}{0pt}%
\pgfpathmoveto{\pgfqpoint{0.474609in}{0.377132in}}%
\pgfpathlineto{\pgfqpoint{0.474609in}{1.414243in}}%
\pgfusepath{stroke}%
\end{pgfscope}%
\begin{pgfscope}%
\definecolor{textcolor}{rgb}{0.150000,0.150000,0.150000}%
\pgfsetstrokecolor{textcolor}%
\pgfsetfillcolor{textcolor}%
\pgftext[x=0.474609in,y=0.261854in,,top]{\color{textcolor}\rmfamily\fontsize{9.000000}{10.800000}\selectfont \(\displaystyle {0}\)}%
\end{pgfscope}%
\begin{pgfscope}%
\pgfpathrectangle{\pgfqpoint{0.474609in}{0.377132in}}{\pgfqpoint{2.881555in}{1.037112in}}%
\pgfusepath{clip}%
\pgfsetroundcap%
\pgfsetroundjoin%
\pgfsetlinewidth{0.803000pt}%
\definecolor{currentstroke}{rgb}{0.800000,0.800000,0.800000}%
\pgfsetstrokecolor{currentstroke}%
\pgfsetdash{}{0pt}%
\pgfpathmoveto{\pgfqpoint{1.245490in}{0.377132in}}%
\pgfpathlineto{\pgfqpoint{1.245490in}{1.414243in}}%
\pgfusepath{stroke}%
\end{pgfscope}%
\begin{pgfscope}%
\definecolor{textcolor}{rgb}{0.150000,0.150000,0.150000}%
\pgfsetstrokecolor{textcolor}%
\pgfsetfillcolor{textcolor}%
\pgftext[x=1.245490in,y=0.261854in,,top]{\color{textcolor}\rmfamily\fontsize{9.000000}{10.800000}\selectfont \(\displaystyle {200000}\)}%
\end{pgfscope}%
\begin{pgfscope}%
\pgfpathrectangle{\pgfqpoint{0.474609in}{0.377132in}}{\pgfqpoint{2.881555in}{1.037112in}}%
\pgfusepath{clip}%
\pgfsetroundcap%
\pgfsetroundjoin%
\pgfsetlinewidth{0.803000pt}%
\definecolor{currentstroke}{rgb}{0.800000,0.800000,0.800000}%
\pgfsetstrokecolor{currentstroke}%
\pgfsetdash{}{0pt}%
\pgfpathmoveto{\pgfqpoint{2.016370in}{0.377132in}}%
\pgfpathlineto{\pgfqpoint{2.016370in}{1.414243in}}%
\pgfusepath{stroke}%
\end{pgfscope}%
\begin{pgfscope}%
\definecolor{textcolor}{rgb}{0.150000,0.150000,0.150000}%
\pgfsetstrokecolor{textcolor}%
\pgfsetfillcolor{textcolor}%
\pgftext[x=2.016370in,y=0.261854in,,top]{\color{textcolor}\rmfamily\fontsize{9.000000}{10.800000}\selectfont \(\displaystyle {400000}\)}%
\end{pgfscope}%
\begin{pgfscope}%
\pgfpathrectangle{\pgfqpoint{0.474609in}{0.377132in}}{\pgfqpoint{2.881555in}{1.037112in}}%
\pgfusepath{clip}%
\pgfsetroundcap%
\pgfsetroundjoin%
\pgfsetlinewidth{0.803000pt}%
\definecolor{currentstroke}{rgb}{0.800000,0.800000,0.800000}%
\pgfsetstrokecolor{currentstroke}%
\pgfsetdash{}{0pt}%
\pgfpathmoveto{\pgfqpoint{2.787251in}{0.377132in}}%
\pgfpathlineto{\pgfqpoint{2.787251in}{1.414243in}}%
\pgfusepath{stroke}%
\end{pgfscope}%
\begin{pgfscope}%
\definecolor{textcolor}{rgb}{0.150000,0.150000,0.150000}%
\pgfsetstrokecolor{textcolor}%
\pgfsetfillcolor{textcolor}%
\pgftext[x=2.787251in,y=0.261854in,,top]{\color{textcolor}\rmfamily\fontsize{9.000000}{10.800000}\selectfont \(\displaystyle {600000}\)}%
\end{pgfscope}%
\begin{pgfscope}%
\definecolor{textcolor}{rgb}{0.150000,0.150000,0.150000}%
\pgfsetstrokecolor{textcolor}%
\pgfsetfillcolor{textcolor}%
\pgftext[x=1.915387in,y=0.085327in,,top]{\color{textcolor}\rmfamily\fontsize{9.000000}{10.800000}\selectfont time \(\displaystyle k\)}%
\end{pgfscope}%
\begin{pgfscope}%
\pgfpathrectangle{\pgfqpoint{0.474609in}{0.377132in}}{\pgfqpoint{2.881555in}{1.037112in}}%
\pgfusepath{clip}%
\pgfsetroundcap%
\pgfsetroundjoin%
\pgfsetlinewidth{0.803000pt}%
\definecolor{currentstroke}{rgb}{0.800000,0.800000,0.800000}%
\pgfsetstrokecolor{currentstroke}%
\pgfsetdash{}{0pt}%
\pgfpathmoveto{\pgfqpoint{0.474609in}{0.399238in}}%
\pgfpathlineto{\pgfqpoint{3.356164in}{0.399238in}}%
\pgfusepath{stroke}%
\end{pgfscope}%
\begin{pgfscope}%
\definecolor{textcolor}{rgb}{0.150000,0.150000,0.150000}%
\pgfsetstrokecolor{textcolor}%
\pgfsetfillcolor{textcolor}%
\pgftext[x=0.291206in, y=0.351753in, left, base]{\color{textcolor}\rmfamily\fontsize{9.000000}{10.800000}\selectfont \(\displaystyle {0}\)}%
\end{pgfscope}%
\begin{pgfscope}%
\pgfpathrectangle{\pgfqpoint{0.474609in}{0.377132in}}{\pgfqpoint{2.881555in}{1.037112in}}%
\pgfusepath{clip}%
\pgfsetroundcap%
\pgfsetroundjoin%
\pgfsetlinewidth{0.803000pt}%
\definecolor{currentstroke}{rgb}{0.800000,0.800000,0.800000}%
\pgfsetstrokecolor{currentstroke}%
\pgfsetdash{}{0pt}%
\pgfpathmoveto{\pgfqpoint{0.474609in}{0.849273in}}%
\pgfpathlineto{\pgfqpoint{3.356164in}{0.849273in}}%
\pgfusepath{stroke}%
\end{pgfscope}%
\begin{pgfscope}%
\definecolor{textcolor}{rgb}{0.150000,0.150000,0.150000}%
\pgfsetstrokecolor{textcolor}%
\pgfsetfillcolor{textcolor}%
\pgftext[x=0.223081in, y=0.801787in, left, base]{\color{textcolor}\rmfamily\fontsize{9.000000}{10.800000}\selectfont \(\displaystyle {20}\)}%
\end{pgfscope}%
\begin{pgfscope}%
\pgfpathrectangle{\pgfqpoint{0.474609in}{0.377132in}}{\pgfqpoint{2.881555in}{1.037112in}}%
\pgfusepath{clip}%
\pgfsetroundcap%
\pgfsetroundjoin%
\pgfsetlinewidth{0.803000pt}%
\definecolor{currentstroke}{rgb}{0.800000,0.800000,0.800000}%
\pgfsetstrokecolor{currentstroke}%
\pgfsetdash{}{0pt}%
\pgfpathmoveto{\pgfqpoint{0.474609in}{1.299307in}}%
\pgfpathlineto{\pgfqpoint{3.356164in}{1.299307in}}%
\pgfusepath{stroke}%
\end{pgfscope}%
\begin{pgfscope}%
\definecolor{textcolor}{rgb}{0.150000,0.150000,0.150000}%
\pgfsetstrokecolor{textcolor}%
\pgfsetfillcolor{textcolor}%
\pgftext[x=0.223081in, y=1.251822in, left, base]{\color{textcolor}\rmfamily\fontsize{9.000000}{10.800000}\selectfont \(\displaystyle {40}\)}%
\end{pgfscope}%
\begin{pgfscope}%
\definecolor{textcolor}{rgb}{0.150000,0.150000,0.150000}%
\pgfsetstrokecolor{textcolor}%
\pgfsetfillcolor{textcolor}%
\pgftext[x=0.167526in,y=0.895687in,,bottom,rotate=90.000000]{\color{textcolor}\rmfamily\fontsize{9.000000}{10.800000}\selectfont observed cost \(\displaystyle \hat J(k)\)}%
\end{pgfscope}%
\begin{pgfscope}%
\pgfsys@transformshift{0.473333in}{0.422325in}%
\pgftext[left,bottom]{\includegraphics[interpolate=true,width=2.753333in,height=0.956667in]{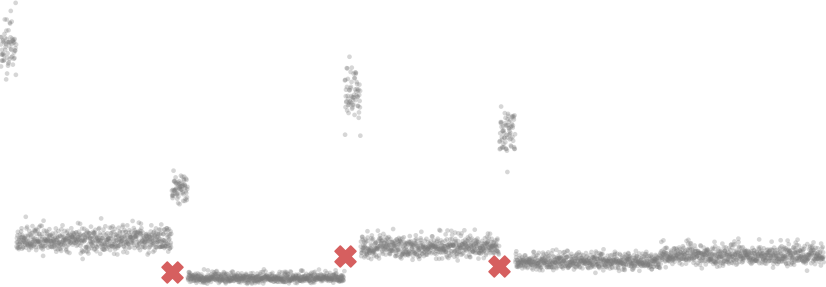}}%
\end{pgfscope}%
\begin{pgfscope}%
\pgfpathrectangle{\pgfqpoint{0.474609in}{0.377132in}}{\pgfqpoint{2.881555in}{1.037112in}}%
\pgfusepath{clip}%
\pgfsetbuttcap%
\pgfsetroundjoin%
\pgfsetlinewidth{1.003750pt}%
\definecolor{currentstroke}{rgb}{0.000000,0.000000,0.000000}%
\pgfsetstrokecolor{currentstroke}%
\pgfsetdash{{6.400000pt}{1.600000pt}{1.000000pt}{1.600000pt}}{0.000000pt}%
\pgfpathmoveto{\pgfqpoint{1.045643in}{0.377132in}}%
\pgfpathlineto{\pgfqpoint{1.045643in}{1.414243in}}%
\pgfusepath{stroke}%
\end{pgfscope}%
\begin{pgfscope}%
\pgfpathrectangle{\pgfqpoint{0.474609in}{0.377132in}}{\pgfqpoint{2.881555in}{1.037112in}}%
\pgfusepath{clip}%
\pgfsetbuttcap%
\pgfsetroundjoin%
\pgfsetlinewidth{1.003750pt}%
\definecolor{currentstroke}{rgb}{0.000000,0.000000,0.000000}%
\pgfsetstrokecolor{currentstroke}%
\pgfsetdash{{6.400000pt}{1.600000pt}{1.000000pt}{1.600000pt}}{0.000000pt}%
\pgfpathmoveto{\pgfqpoint{1.620527in}{0.377132in}}%
\pgfpathlineto{\pgfqpoint{1.620527in}{1.414243in}}%
\pgfusepath{stroke}%
\end{pgfscope}%
\begin{pgfscope}%
\pgfpathrectangle{\pgfqpoint{0.474609in}{0.377132in}}{\pgfqpoint{2.881555in}{1.037112in}}%
\pgfusepath{clip}%
\pgfsetbuttcap%
\pgfsetroundjoin%
\pgfsetlinewidth{1.003750pt}%
\definecolor{currentstroke}{rgb}{0.000000,0.000000,0.000000}%
\pgfsetstrokecolor{currentstroke}%
\pgfsetdash{{6.400000pt}{1.600000pt}{1.000000pt}{1.600000pt}}{0.000000pt}%
\pgfpathmoveto{\pgfqpoint{2.133741in}{0.377132in}}%
\pgfpathlineto{\pgfqpoint{2.133741in}{1.414243in}}%
\pgfusepath{stroke}%
\end{pgfscope}%
\begin{pgfscope}%
\pgfpathrectangle{\pgfqpoint{0.474609in}{0.377132in}}{\pgfqpoint{2.881555in}{1.037112in}}%
\pgfusepath{clip}%
\pgfsetbuttcap%
\pgfsetroundjoin%
\pgfsetlinewidth{1.003750pt}%
\definecolor{currentstroke}{rgb}{0.000000,0.000000,0.000000}%
\pgfsetstrokecolor{currentstroke}%
\pgfsetdash{{6.400000pt}{1.600000pt}{1.000000pt}{1.600000pt}}{0.000000pt}%
\pgfpathmoveto{\pgfqpoint{2.677019in}{0.377132in}}%
\pgfpathlineto{\pgfqpoint{2.677019in}{1.414243in}}%
\pgfusepath{stroke}%
\end{pgfscope}%
\begin{pgfscope}%
\pgfpathrectangle{\pgfqpoint{0.474609in}{0.377132in}}{\pgfqpoint{2.881555in}{1.037112in}}%
\pgfusepath{clip}%
\pgfsetbuttcap%
\pgfsetroundjoin%
\pgfsetlinewidth{3.011250pt}%
\definecolor{currentstroke}{rgb}{0.282353,0.470588,0.815686}%
\pgfsetstrokecolor{currentstroke}%
\pgfsetdash{}{0pt}%
\pgfpathmoveto{\pgfqpoint{0.474613in}{0.502754in}}%
\pgfpathlineto{\pgfqpoint{0.474613in}{0.502754in}}%
\pgfusepath{stroke}%
\end{pgfscope}%
\begin{pgfscope}%
\pgfsys@transformshift{0.526667in}{0.478991in}%
\pgftext[left,bottom]{\includegraphics[interpolate=true,width=0.523333in,height=0.253333in]{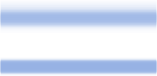}}%
\end{pgfscope}%
\begin{pgfscope}%
\pgfpathrectangle{\pgfqpoint{0.474609in}{0.377132in}}{\pgfqpoint{2.881555in}{1.037112in}}%
\pgfusepath{clip}%
\pgfsetbuttcap%
\pgfsetroundjoin%
\pgfsetlinewidth{2.007500pt}%
\definecolor{currentstroke}{rgb}{0.415686,0.800000,0.392157}%
\pgfsetstrokecolor{currentstroke}%
\pgfsetdash{{2.000000pt}{3.300000pt}}{0.000000pt}%
\pgfpathmoveto{\pgfqpoint{0.474613in}{0.636590in}}%
\pgfpathlineto{\pgfqpoint{1.045643in}{0.636590in}}%
\pgfusepath{stroke}%
\end{pgfscope}%
\begin{pgfscope}%
\pgfsys@transformshift{1.100000in}{0.402325in}%
\pgftext[left,bottom]{\includegraphics[interpolate=true,width=0.526667in,height=0.126667in]{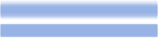}}%
\end{pgfscope}%
\begin{pgfscope}%
\pgfpathrectangle{\pgfqpoint{0.474609in}{0.377132in}}{\pgfqpoint{2.881555in}{1.037112in}}%
\pgfusepath{clip}%
\pgfsetbuttcap%
\pgfsetroundjoin%
\pgfsetlinewidth{2.007500pt}%
\definecolor{currentstroke}{rgb}{0.415686,0.800000,0.392157}%
\pgfsetstrokecolor{currentstroke}%
\pgfsetdash{{2.000000pt}{3.300000pt}}{0.000000pt}%
\pgfpathmoveto{\pgfqpoint{1.045643in}{0.477496in}}%
\pgfpathlineto{\pgfqpoint{1.622069in}{0.477496in}}%
\pgfusepath{stroke}%
\end{pgfscope}%
\begin{pgfscope}%
\pgfsys@transformshift{1.676667in}{0.468991in}%
\pgftext[left,bottom]{\includegraphics[interpolate=true,width=0.463333in,height=0.280000in]{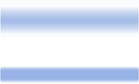}}%
\end{pgfscope}%
\begin{pgfscope}%
\pgfpathrectangle{\pgfqpoint{0.474609in}{0.377132in}}{\pgfqpoint{2.881555in}{1.037112in}}%
\pgfusepath{clip}%
\pgfsetbuttcap%
\pgfsetroundjoin%
\pgfsetlinewidth{2.007500pt}%
\definecolor{currentstroke}{rgb}{0.415686,0.800000,0.392157}%
\pgfsetstrokecolor{currentstroke}%
\pgfsetdash{{2.000000pt}{3.300000pt}}{0.000000pt}%
\pgfpathmoveto{\pgfqpoint{1.622069in}{0.605111in}}%
\pgfpathlineto{\pgfqpoint{2.137595in}{0.605111in}}%
\pgfusepath{stroke}%
\end{pgfscope}%
\begin{pgfscope}%
\pgfsys@transformshift{2.190000in}{0.442325in}%
\pgftext[left,bottom]{\includegraphics[interpolate=true,width=1.033333in,height=0.210000in]{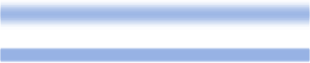}}%
\end{pgfscope}%
\begin{pgfscope}%
\pgfpathrectangle{\pgfqpoint{0.474609in}{0.377132in}}{\pgfqpoint{2.881555in}{1.037112in}}%
\pgfusepath{clip}%
\pgfsetbuttcap%
\pgfsetroundjoin%
\pgfsetlinewidth{2.007500pt}%
\definecolor{currentstroke}{rgb}{0.415686,0.800000,0.392157}%
\pgfsetstrokecolor{currentstroke}%
\pgfsetdash{{2.000000pt}{3.300000pt}}{0.000000pt}%
\pgfpathmoveto{\pgfqpoint{2.137595in}{0.536289in}}%
\pgfpathlineto{\pgfqpoint{3.218948in}{0.536289in}}%
\pgfusepath{stroke}%
\end{pgfscope}%
\begin{pgfscope}%
\pgfpathrectangle{\pgfqpoint{0.474609in}{0.377132in}}{\pgfqpoint{2.881555in}{1.037112in}}%
\pgfusepath{clip}%
\pgfsetbuttcap%
\pgfsetroundjoin%
\pgfsetlinewidth{2.007500pt}%
\definecolor{currentstroke}{rgb}{0.584314,0.423529,0.705882}%
\pgfsetstrokecolor{currentstroke}%
\pgfsetdash{{2.000000pt}{3.300000pt}}{0.000000pt}%
\pgfpathmoveto{\pgfqpoint{0.474613in}{1.207972in}}%
\pgfpathlineto{\pgfqpoint{1.045643in}{1.207972in}}%
\pgfusepath{stroke}%
\end{pgfscope}%
\begin{pgfscope}%
\pgfpathrectangle{\pgfqpoint{0.474609in}{0.377132in}}{\pgfqpoint{2.881555in}{1.037112in}}%
\pgfusepath{clip}%
\pgfsetbuttcap%
\pgfsetroundjoin%
\pgfsetlinewidth{2.007500pt}%
\definecolor{currentstroke}{rgb}{0.584314,0.423529,0.705882}%
\pgfsetstrokecolor{currentstroke}%
\pgfsetdash{{2.000000pt}{3.300000pt}}{0.000000pt}%
\pgfpathmoveto{\pgfqpoint{1.045643in}{0.730477in}}%
\pgfpathlineto{\pgfqpoint{1.620527in}{0.730477in}}%
\pgfusepath{stroke}%
\end{pgfscope}%
\begin{pgfscope}%
\pgfpathrectangle{\pgfqpoint{0.474609in}{0.377132in}}{\pgfqpoint{2.881555in}{1.037112in}}%
\pgfusepath{clip}%
\pgfsetbuttcap%
\pgfsetroundjoin%
\pgfsetlinewidth{2.007500pt}%
\definecolor{currentstroke}{rgb}{0.584314,0.423529,0.705882}%
\pgfsetstrokecolor{currentstroke}%
\pgfsetdash{{2.000000pt}{3.300000pt}}{0.000000pt}%
\pgfpathmoveto{\pgfqpoint{1.620527in}{1.054884in}}%
\pgfpathlineto{\pgfqpoint{1.622069in}{1.054884in}}%
\pgfusepath{stroke}%
\end{pgfscope}%
\begin{pgfscope}%
\pgfpathrectangle{\pgfqpoint{0.474609in}{0.377132in}}{\pgfqpoint{2.881555in}{1.037112in}}%
\pgfusepath{clip}%
\pgfsetbuttcap%
\pgfsetroundjoin%
\pgfsetlinewidth{2.007500pt}%
\definecolor{currentstroke}{rgb}{0.584314,0.423529,0.705882}%
\pgfsetstrokecolor{currentstroke}%
\pgfsetdash{{2.000000pt}{3.300000pt}}{0.000000pt}%
\pgfpathmoveto{\pgfqpoint{1.622069in}{1.054884in}}%
\pgfpathlineto{\pgfqpoint{2.133741in}{1.054884in}}%
\pgfusepath{stroke}%
\end{pgfscope}%
\begin{pgfscope}%
\pgfpathrectangle{\pgfqpoint{0.474609in}{0.377132in}}{\pgfqpoint{2.881555in}{1.037112in}}%
\pgfusepath{clip}%
\pgfsetbuttcap%
\pgfsetroundjoin%
\pgfsetlinewidth{2.007500pt}%
\definecolor{currentstroke}{rgb}{0.584314,0.423529,0.705882}%
\pgfsetstrokecolor{currentstroke}%
\pgfsetdash{{2.000000pt}{3.300000pt}}{0.000000pt}%
\pgfpathmoveto{\pgfqpoint{2.133741in}{0.926503in}}%
\pgfpathlineto{\pgfqpoint{2.137595in}{0.926503in}}%
\pgfusepath{stroke}%
\end{pgfscope}%
\begin{pgfscope}%
\pgfpathrectangle{\pgfqpoint{0.474609in}{0.377132in}}{\pgfqpoint{2.881555in}{1.037112in}}%
\pgfusepath{clip}%
\pgfsetbuttcap%
\pgfsetroundjoin%
\pgfsetlinewidth{2.007500pt}%
\definecolor{currentstroke}{rgb}{0.584314,0.423529,0.705882}%
\pgfsetstrokecolor{currentstroke}%
\pgfsetdash{{2.000000pt}{3.300000pt}}{0.000000pt}%
\pgfpathmoveto{\pgfqpoint{2.137595in}{0.926503in}}%
\pgfpathlineto{\pgfqpoint{2.677019in}{0.926503in}}%
\pgfusepath{stroke}%
\end{pgfscope}%
\begin{pgfscope}%
\pgfpathrectangle{\pgfqpoint{0.474609in}{0.377132in}}{\pgfqpoint{2.881555in}{1.037112in}}%
\pgfusepath{clip}%
\pgfsetbuttcap%
\pgfsetroundjoin%
\pgfsetlinewidth{2.007500pt}%
\definecolor{currentstroke}{rgb}{0.584314,0.423529,0.705882}%
\pgfsetstrokecolor{currentstroke}%
\pgfsetdash{{2.000000pt}{3.300000pt}}{0.000000pt}%
\pgfpathmoveto{\pgfqpoint{2.677019in}{0.978298in}}%
\pgfpathlineto{\pgfqpoint{3.218948in}{0.978298in}}%
\pgfusepath{stroke}%
\end{pgfscope}%
\begin{pgfscope}%
\pgfsetrectcap%
\pgfsetmiterjoin%
\pgfsetlinewidth{1.003750pt}%
\definecolor{currentstroke}{rgb}{0.800000,0.800000,0.800000}%
\pgfsetstrokecolor{currentstroke}%
\pgfsetdash{}{0pt}%
\pgfpathmoveto{\pgfqpoint{0.474609in}{0.377132in}}%
\pgfpathlineto{\pgfqpoint{0.474609in}{1.414243in}}%
\pgfusepath{stroke}%
\end{pgfscope}%
\begin{pgfscope}%
\pgfsetrectcap%
\pgfsetmiterjoin%
\pgfsetlinewidth{1.003750pt}%
\definecolor{currentstroke}{rgb}{0.800000,0.800000,0.800000}%
\pgfsetstrokecolor{currentstroke}%
\pgfsetdash{}{0pt}%
\pgfpathmoveto{\pgfqpoint{0.474609in}{0.377132in}}%
\pgfpathlineto{\pgfqpoint{3.356164in}{0.377132in}}%
\pgfusepath{stroke}%
\end{pgfscope}%
\begin{pgfscope}%
\pgfsetbuttcap%
\pgfsetmiterjoin%
\definecolor{currentfill}{rgb}{1.000000,1.000000,1.000000}%
\pgfsetfillcolor{currentfill}%
\pgfsetfillopacity{0.800000}%
\pgfsetlinewidth{0.803000pt}%
\definecolor{currentstroke}{rgb}{0.800000,0.800000,0.800000}%
\pgfsetstrokecolor{currentstroke}%
\pgfsetstrokeopacity{0.800000}%
\pgfsetdash{}{0pt}%
\pgfpathmoveto{\pgfqpoint{0.586593in}{1.476743in}}%
\pgfpathlineto{\pgfqpoint{3.244181in}{1.476743in}}%
\pgfpathquadraticcurveto{\pgfqpoint{3.269181in}{1.476743in}}{\pgfqpoint{3.269181in}{1.501743in}}%
\pgfpathlineto{\pgfqpoint{3.269181in}{1.884026in}}%
\pgfpathquadraticcurveto{\pgfqpoint{3.269181in}{1.909026in}}{\pgfqpoint{3.244181in}{1.909026in}}%
\pgfpathlineto{\pgfqpoint{0.586593in}{1.909026in}}%
\pgfpathquadraticcurveto{\pgfqpoint{0.561593in}{1.909026in}}{\pgfqpoint{0.561593in}{1.884026in}}%
\pgfpathlineto{\pgfqpoint{0.561593in}{1.501743in}}%
\pgfpathquadraticcurveto{\pgfqpoint{0.561593in}{1.476743in}}{\pgfqpoint{0.586593in}{1.476743in}}%
\pgfpathlineto{\pgfqpoint{0.586593in}{1.476743in}}%
\pgfpathclose%
\pgfusepath{stroke,fill}%
\end{pgfscope}%
\begin{pgfscope}%
\pgfsetbuttcap%
\pgfsetroundjoin%
\pgfsetlinewidth{1.003750pt}%
\definecolor{currentstroke}{rgb}{0.000000,0.000000,0.000000}%
\pgfsetstrokecolor{currentstroke}%
\pgfsetdash{{6.400000pt}{1.600000pt}{1.000000pt}{1.600000pt}}{0.000000pt}%
\pgfpathmoveto{\pgfqpoint{0.611593in}{1.807806in}}%
\pgfpathlineto{\pgfqpoint{0.736593in}{1.807806in}}%
\pgfpathlineto{\pgfqpoint{0.861593in}{1.807806in}}%
\pgfusepath{stroke}%
\end{pgfscope}%
\begin{pgfscope}%
\definecolor{textcolor}{rgb}{0.150000,0.150000,0.150000}%
\pgfsetstrokecolor{textcolor}%
\pgfsetfillcolor{textcolor}%
\pgftext[x=0.961593in,y=1.764056in,left,base]{\color{textcolor}\rmfamily\fontsize{9.000000}{10.800000}\selectfont \(\displaystyle T_i\)}%
\end{pgfscope}%
\begin{pgfscope}%
\pgfsetbuttcap%
\pgfsetmiterjoin%
\definecolor{currentfill}{rgb}{0.839216,0.372549,0.372549}%
\pgfsetfillcolor{currentfill}%
\pgfsetlinewidth{1.003750pt}%
\definecolor{currentstroke}{rgb}{0.839216,0.372549,0.372549}%
\pgfsetstrokecolor{currentstroke}%
\pgfsetdash{}{0pt}%
\pgfsys@defobject{currentmarker}{\pgfqpoint{-0.033333in}{-0.033333in}}{\pgfqpoint{0.033333in}{0.033333in}}{%
\pgfpathmoveto{\pgfqpoint{-0.016667in}{-0.033333in}}%
\pgfpathlineto{\pgfqpoint{0.000000in}{-0.016667in}}%
\pgfpathlineto{\pgfqpoint{0.016667in}{-0.033333in}}%
\pgfpathlineto{\pgfqpoint{0.033333in}{-0.016667in}}%
\pgfpathlineto{\pgfqpoint{0.016667in}{0.000000in}}%
\pgfpathlineto{\pgfqpoint{0.033333in}{0.016667in}}%
\pgfpathlineto{\pgfqpoint{0.016667in}{0.033333in}}%
\pgfpathlineto{\pgfqpoint{0.000000in}{0.016667in}}%
\pgfpathlineto{\pgfqpoint{-0.016667in}{0.033333in}}%
\pgfpathlineto{\pgfqpoint{-0.033333in}{0.016667in}}%
\pgfpathlineto{\pgfqpoint{-0.016667in}{0.000000in}}%
\pgfpathlineto{\pgfqpoint{-0.033333in}{-0.016667in}}%
\pgfpathlineto{\pgfqpoint{-0.016667in}{-0.033333in}}%
\pgfpathclose%
\pgfusepath{stroke,fill}%
}%
\begin{pgfscope}%
\pgfsys@transformshift{0.736593in}{1.623892in}%
\pgfsys@useobject{currentmarker}{}%
\end{pgfscope}%
\end{pgfscope}%
\begin{pgfscope}%
\definecolor{textcolor}{rgb}{0.150000,0.150000,0.150000}%
\pgfsetstrokecolor{textcolor}%
\pgfsetfillcolor{textcolor}%
\pgftext[x=0.961593in,y=1.580142in,left,base]{\color{textcolor}\rmfamily\fontsize{9.000000}{10.800000}\selectfont trigger}%
\end{pgfscope}%
\begin{pgfscope}%
\pgfsetbuttcap%
\pgfsetroundjoin%
\pgfsetlinewidth{2.007500pt}%
\definecolor{currentstroke}{rgb}{0.584314,0.423529,0.705882}%
\pgfsetstrokecolor{currentstroke}%
\pgfsetdash{{2.000000pt}{3.300000pt}}{0.000000pt}%
\pgfpathmoveto{\pgfqpoint{1.655318in}{1.807806in}}%
\pgfpathlineto{\pgfqpoint{1.780318in}{1.807806in}}%
\pgfpathlineto{\pgfqpoint{1.905318in}{1.807806in}}%
\pgfusepath{stroke}%
\end{pgfscope}%
\begin{pgfscope}%
\definecolor{textcolor}{rgb}{0.150000,0.150000,0.150000}%
\pgfsetstrokecolor{textcolor}%
\pgfsetfillcolor{textcolor}%
\pgftext[x=2.005318in,y=1.764056in,left,base]{\color{textcolor}\rmfamily\fontsize{9.000000}{10.800000}\selectfont \(\displaystyle \Delta J(K_0)\)}%
\end{pgfscope}%
\begin{pgfscope}%
\pgfsetbuttcap%
\pgfsetroundjoin%
\pgfsetlinewidth{2.007500pt}%
\definecolor{currentstroke}{rgb}{0.415686,0.800000,0.392157}%
\pgfsetstrokecolor{currentstroke}%
\pgfsetdash{{2.000000pt}{3.300000pt}}{0.000000pt}%
\pgfpathmoveto{\pgfqpoint{1.655318in}{1.602118in}}%
\pgfpathlineto{\pgfqpoint{1.780318in}{1.602118in}}%
\pgfpathlineto{\pgfqpoint{1.905318in}{1.602118in}}%
\pgfusepath{stroke}%
\end{pgfscope}%
\begin{pgfscope}%
\definecolor{textcolor}{rgb}{0.150000,0.150000,0.150000}%
\pgfsetstrokecolor{textcolor}%
\pgfsetfillcolor{textcolor}%
\pgftext[x=2.005318in,y=1.558368in,left,base]{\color{textcolor}\rmfamily\fontsize{9.000000}{10.800000}\selectfont \(\displaystyle J_{\Delta}(\hat N^*)\)}%
\end{pgfscope}%
\begin{pgfscope}%
\pgfsetbuttcap%
\pgfsetroundjoin%
\pgfsetlinewidth{3.011250pt}%
\definecolor{currentstroke}{rgb}{0.282353,0.470588,0.815686}%
\pgfsetstrokecolor{currentstroke}%
\pgfsetdash{}{0pt}%
\pgfpathmoveto{\pgfqpoint{2.631540in}{1.807806in}}%
\pgfpathlineto{\pgfqpoint{2.756540in}{1.807806in}}%
\pgfpathlineto{\pgfqpoint{2.881540in}{1.807806in}}%
\pgfusepath{stroke}%
\end{pgfscope}%
\begin{pgfscope}%
\definecolor{textcolor}{rgb}{0.150000,0.150000,0.150000}%
\pgfsetstrokecolor{textcolor}%
\pgfsetfillcolor{textcolor}%
\pgftext[x=2.981540in,y=1.764056in,left,base]{\color{textcolor}\rmfamily\fontsize{9.000000}{10.800000}\selectfont \(\displaystyle p(\kappa\))}%
\end{pgfscope}%
\end{pgfpicture}%
\makeatother%
\endgroup%